\documentclass[aps,prapplied,amsmath,amssymb,reprint,superscriptaddress]{revtex4-2}

\usepackage[utf8]{inputenc}
\usepackage[T1]{fontenc}
\usepackage{mathptmx}

\usepackage{lineno,hyperref}
\usepackage{graphicx}
\newcommand{\YIG}{Y$_3$Fe$_5$O$_{12}$}
\newcommand{\GIG}{Gd$_3$Fe$_5$O$_{12}$}
\newcommand{\GGG}{Gd$_3$Ga$_5$O$_{12}$}
\newcommand{\TcompG}{\ensuremath{T_{comp,G}}}
\newcommand{\TcompB}{\ensuremath{T_{comp,B}}}

\begin{document}

\title{\texorpdfstring{Magnetic coupling in \YIG/\GIG\ heterostructures}{Magnetic coupling in Y3Fe5O12/Ge3Fe5O12 heterostructures}}

\author{S.~Becker}
	\email{svenbecker@uni-mainz.de}
\affiliation{Institute of Physics, Johannes Gutenberg-University Mainz, Staudingerweg 7, 55128 Mainz, Germany}
\author{Z.~Ren}
	\email{zengyaoren@163.com}
\altaffiliation{S.B. and Z.R. contributed equally to this work}
\affiliation{Institute of Physics, Johannes Gutenberg-University Mainz, Staudingerweg 7, 55128 Mainz, Germany}
\affiliation{Graduate School of Excellence “Materials Science in Mainz” (MAINZ), Staudingerweg 9, 55128 Mainz}
\affiliation{School of Materials Science and Engineering, University of Science and Technology Beijing, Beijing 100083, China}
\author{F.~Fuhrmann}
\affiliation{Institute of Physics, Johannes Gutenberg-University Mainz, Staudingerweg 7, 55128 Mainz, Germany}
\author{A.~Ross}
\affiliation{Unité Mixte de Physique CNRS, Thales, Université Paris-Saclay, 91767 Palaiseau, France}
\affiliation{Institute of Physics, Johannes Gutenberg-University Mainz, Staudingerweg 7, 55128 Mainz, Germany}
\author{S.~Lord}
\affiliation{Institute of Physics, Johannes Gutenberg-University Mainz, Staudingerweg 7, 55128 Mainz, Germany}
\affiliation{Graduate School of Excellence “Materials Science in Mainz” (MAINZ), Staudingerweg 9, 55128 Mainz}
\affiliation{Department of Physics and Astronomy, University of Manchester, Manchester M13 9PL, United Kingdom}
\author{S.~Ding}
\affiliation{Institute of Physics, Johannes Gutenberg-University Mainz, Staudingerweg 7, 55128 Mainz, Germany}
\affiliation{Graduate School of Excellence “Materials Science in Mainz” (MAINZ), Staudingerweg 9, 55128 Mainz}
\affiliation{State Key Laboratory for Mesoscopic Physics, School of Physics, Peking University, Beijing 100871, China}
\author{R.~Wu}
\affiliation{Institute of Physics, Johannes Gutenberg-University Mainz, Staudingerweg 7, 55128 Mainz, Germany}
\author{J.~Yang}
\affiliation{State Key Laboratory for Mesoscopic Physics, School of Physics, Peking University, Beijing 100871, China}
\author{J.~Miao}
\affiliation{School of Materials Science and Engineering, University of Science and Technology Beijing, Beijing 100083, China}
\author{M.~Kl\"aui}
\affiliation{Institute of Physics, Johannes Gutenberg-University Mainz, Staudingerweg 7, 55128 Mainz, Germany}
\affiliation{Graduate School of Excellence “Materials Science in Mainz” (MAINZ), Staudingerweg 9, 55128 Mainz}
\affiliation{Center for Quantum Spintronics, Norwegian University of Science and Technology, 7491\,Trondheim, Norway}
\author{G.~Jakob}
	\email{jakob@uni-mainz.de}
\affiliation{Institute of Physics, Johannes Gutenberg-University Mainz, Staudingerweg 7, 55128 Mainz, Germany}
\affiliation{Graduate School of Excellence “Materials Science in Mainz” (MAINZ), Staudingerweg 9, 55128 Mainz}

\date{\today}
\begin{abstract}
Ferrimagnetic \YIG\ (YIG) is the prototypical material for studying magnonic properties due to its exceptionally low damping. By substituting the yttrium with other rare earth elements that have a net magnetic moment, we can introduce an additional spin degree of freedom. Here, we study the magnetic coupling in epitaxial \YIG/\GIG\ (YIG/GIG) heterostructures grown by pulsed laser deposition. From bulk sensitive magnetometry and surface sensitive spin Seebeck effect (SSE) and spin Hall magnetoresistance (SMR) measurements, we determine the alignment of the heterostructure magnetization through temperature and external magnetic field. The ferromagnetic coupling between the Fe sublattices of YIG and GIG dominates the overall behavior of the heterostructures. Due to the temperature dependent gadolinium moment, a magnetic compensation point of the total bilayer system can be identified. This compensation point shifts to lower temperatures with increasing thickness of YIG due the parallel alignment of the iron moments. We show that we can control the magnetic properties of the heterostructures by tuning the thickness of the individual layers, opening up a large playground for magnonic devices based on coupled magnetic insulators. These devices could potentially control the magnon transport analogously to electron transport in giant magnetoresistive devices. 
\end{abstract}
\maketitle

\section{Introduction}
A major challenge in information technology is solving the issue of Joule heating due to charge currents. One approach is to move away from electron-based to magnon-mediated information transport and processing \cite{Chumak2015}. This requires the developments of new logic devices, such as magnon valves that allow for the manipulation of spin currents \cite{Cramer2018}. Material candidates require an insulating character, magnetic ordering and low magnetic damping. 
One promising candidate is ferrimagnetic \YIG\ (YIG = yttrium iron garnet), which is a well-known material in magnetism as it shows ultra-low magnetic damping and low magnetic anisotropy. 
The net ferrimagnetic moment originates from an antiparallel alignment of Fe$^{3+}$ moments on different crystallographic sites. Each minimal unit cell consists of 12 trivalent Fe$^{3+}$ ions that are tetrahedrally coordinated with oxygen atoms ($d$ sites) and 8 trivalent Fe$^{3+}$ ions that are octahedrally coordinated ($a$ sites). The dominant coupling is antiferromagnetic between iron atoms on minority $a$ and majority $d$ sites. 
By substituting Y$^{3+}$ with Gd$^{3+}$, an additional moment appears aligned antiparallel to the $d$-site Fe atoms \cite{Harris1963, Yamagishi2005}. 
Due to the strong temperature dependence of the Gd net magnetic moment, \GIG\ (GIG) shows a magnetization compensation at temperature $T\approx295$\,K \cite{Pauthenet1958}.

For low power information processing, magnons (the quanta of magnetic excitation) are exciting candidates.
The magnon spectra of YIG has been the subject of both experimental and theoretical investigation \cite{Cherepanov1993,Princep2017}.
In heterostructures, magnon-magnon coupling and magnetic coupling play a decisive role for the propagation of magnons \cite{Cramer2018,Wu2018,Guo2018,Fan2020,Guo2020,Li2020,Kumar2021}. 
Here, we investigate the coupling between two different iron based garnets YIG and GIG as candidates for an all insulator magnon spin valve.  
Both YIG and GIG are important ferrimagnetic insulators that can be grown epitaxially on isostructural but paramagnetic \GGG\ (GGG) substrates. 
Recently, magnetic coupling of YIG to an ultrathin GIG layer was reported \cite{Gomez2018,Kumar2021}, where GIG is formed by an interdiffusion process while preparing YIG on GGG. 
In our work, we realize controlled growth of epitaxial YIG/GIG heterostructures on GGG substrates, where the individual layers have sufficient magnetic moment to be detected by magnetometry. 
The bilayers show high crystalline quality and a magnetic compensation of the entire bilayer, which shifts to lower temperature with increasing thickness of YIG, indicating interlayer magnetic coupling of YIG and GIG. The alignment of the magnetization of bilayers with temperature and field can be characterized by SQUID magnetometry, which measures the sum signal of all the layers. To identify the magnetization direction, we utilize the surface sensitive spin Hall magnetoresistance (SMR), which has proven to be a suitable tool to investigate the magnetic properties of magnetic insulators \cite{Nakayama2013,Ganzhorn2016,Dong2017,Fayaz2019}. 
We further conduct spin Seebeck (SSE) measurements, which have previously been used to investigate pure GIG samples \cite{Fayaz2019,Uchida2010,Kehlberger2015,Guo2016,Cramer2017,Gepraegs2016}. SSE measurements show dominating sensitivity to the top layer of the heterostructure. 
Our results demonstrate that the respective Fe sublattices of YIG and GIG are ferromagnetically coupled across the interface. This allows for an unprecedented tunability of the magnetic properties of the bilayer system by choosing the relative thickness of the YIG and GIG layers. 

\section{Experimental details}
\YIG\ (YIG) and \GIG (GIG) are deposited on (001)-oriented \GGG\ (GGG) substrates by pulsed laser deposition (PLD) in an ultrahigh vacuum chamber with a base pressure lower than $2\times10^{-8}$\,mbar. For ablation, a KrF excimer laser (248\,nm wavelength) with a nominal energy of 130\,mJ per pulse is used at a pulse frequency of 10\,Hz. The films are grown under a stable atmosphere of 0.026\,mbar of O$_2$ at $475^\circ$C substrate temperature. After deposition, the films are subsequently cooled down to room temperature at a rate of $-25$\,K/min. The crystalline structure of the films was determined by x-ray diffraction (XRD). The magnetic moment was measured by using a superconducting quantum interference device magnetometer (SQUID, Quantum Design MPMS II). For spin Seebeck effect (SSE) measurements, the samples are covered with a continuous layer of 4\,nm Pt deposited by magneton sputtering. The measurements are performed in the longitudinal geometry, where the heat gradient is perpendicular to the sample surface \cite{Uchida2010}. For spin Hall magnetoresistance (SMR) measurements, a 4\,nm thick Pt bar of around 0.3\,mm width is defined on the surface along the crystallographic axes. The Pt is deposited \textit{ex-situ} using magneton sputtering in an Ar atmosphere through a shadow mask.

\section{Results}
\begin{figure}
	\includegraphics[width=8.5cm]{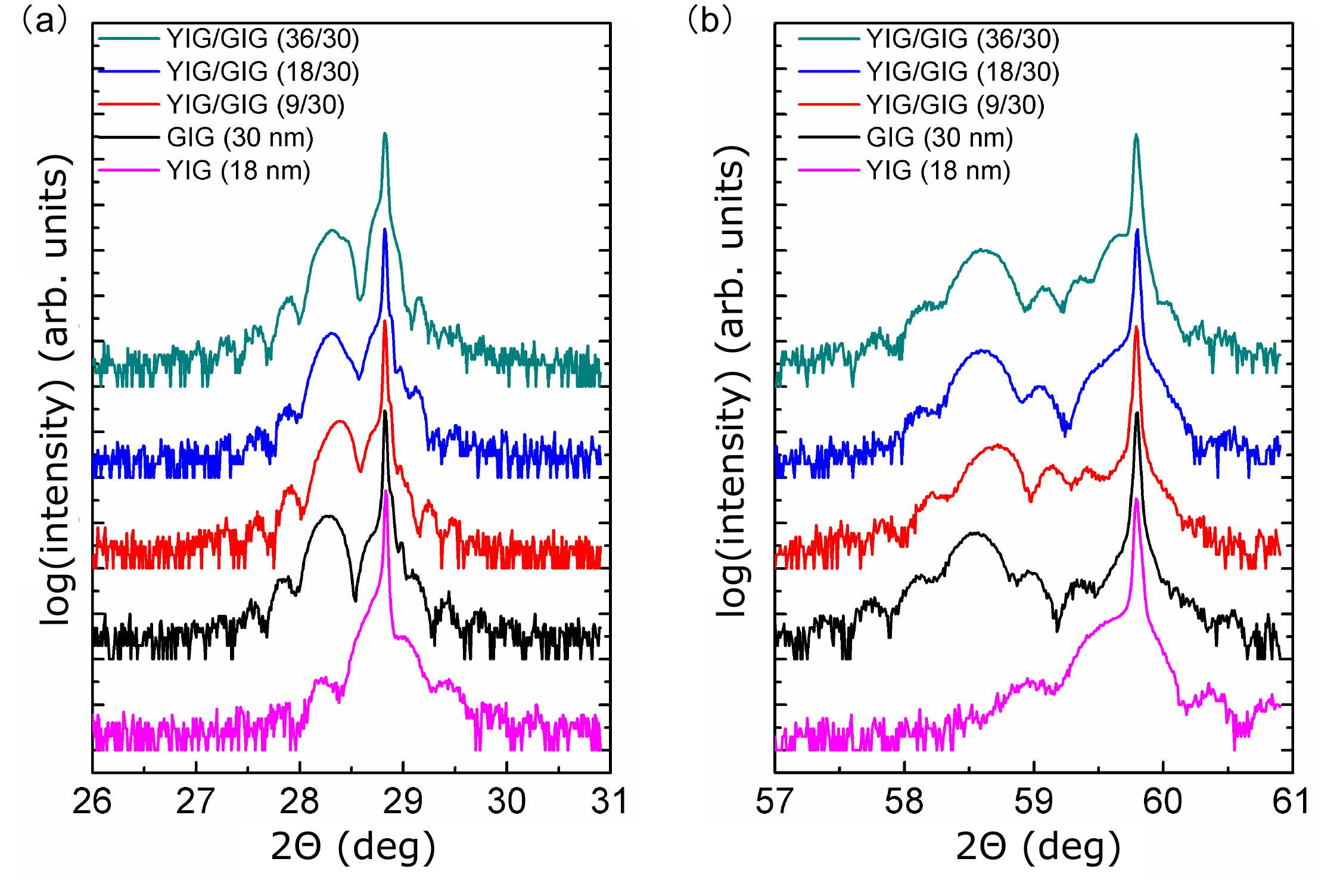}
	\caption{\label{fig:XRD}Out-of-plane $2\Theta/\omega$ measurements of YIG, GIG and YIG/GIG bilayer films near the (004) peaks (a) and (008) peaks (b) shown in a logarithmic intensity scale. The thicknesses of the individual layers are detailed in nanometers. }
\end{figure}

Fig.\ \ref{fig:XRD}(a) and \ref{fig:XRD}(b) present the XRD patterns of GGG/YIG, GGG/GIG, and GGG/YIG/GIG bilayer films measured with the scattering vector normal to the (001) oriented cubic substrate. As the films grow coherently on the substrate surface, the $a$ and $b$ axes are equally strained and the (004) peaks and (008) peaks, indicating the length of the $c$-axis, are evident in the corresponding high resolution XRD patterns, respectively. Near the respective (004) (Fig.\ \ref{fig:XRD}(a)) and (008) (Fig.\ \ref{fig:XRD}(b)) diffraction peaks, the XRD patterns show Laue oscillations, indicating a smooth surface and interface. While the YIG reflex is partly shadowed by the substrate peak, we can clearly identify the reflex of the GIG top layer. From the GIG peak position, we determine the out-of-plane lattice parameter to around $c = 1.26$\,nm, independent of the thickness of the underlying YIG layer, indicating strained growth for every sample. The absence of a structural alteration of the top GIG layer indicates that the YIG interlayer does not influence the GIG growth. It is therefore expected, that the magnetic properties of the GIG layer are similarly unaffected. 
The rocking curve of each reflex is around $\Delta\omega=0.04^\circ$ further showing well aligned unit cells for every bilayer. Together, these XRD measurements indicate the high-quality growth of YIG/GIG heterostructures. 

\begin{figure}
	\includegraphics[width=8.5cm]{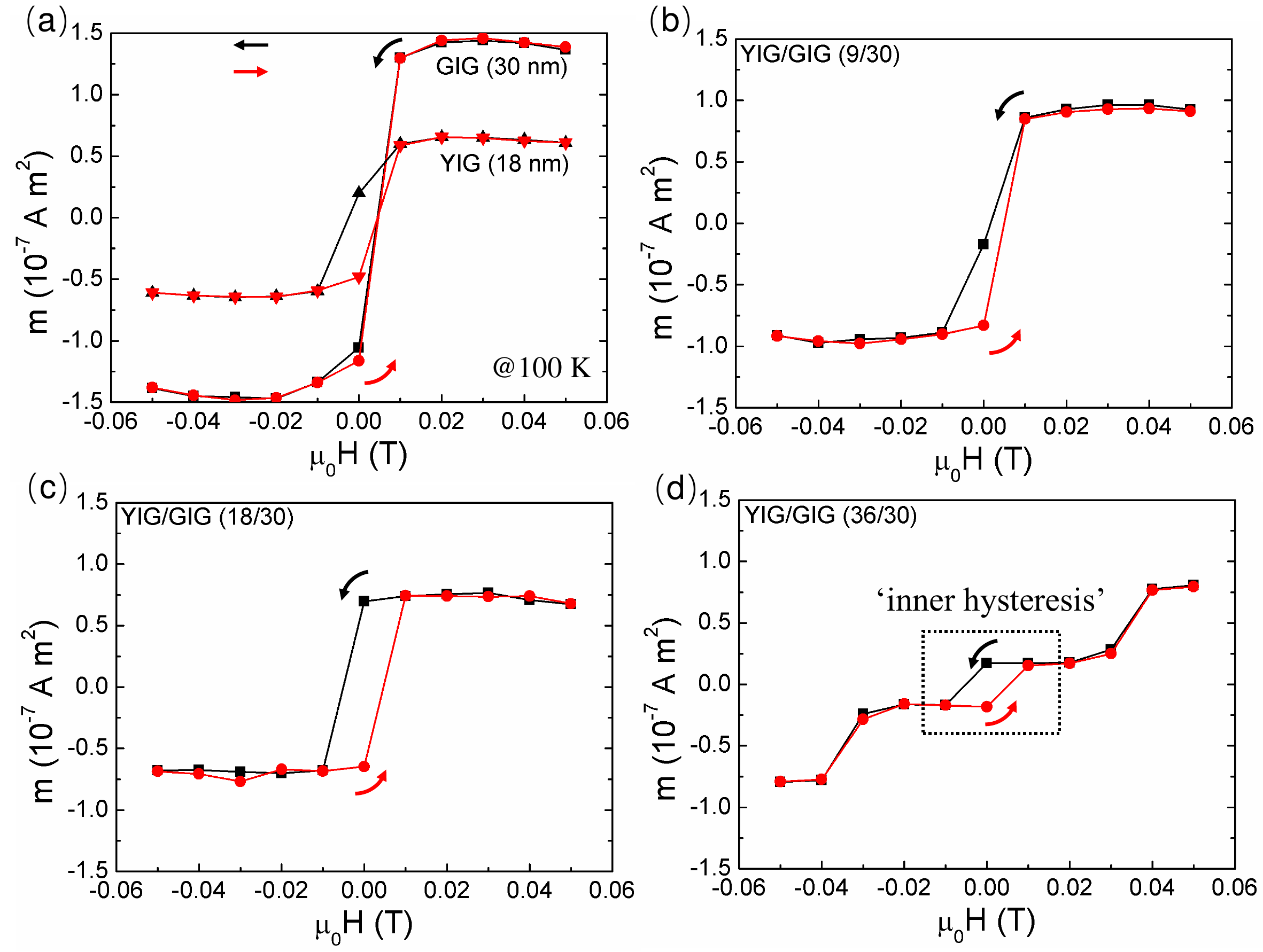}
	\caption{\label{fig:SQUID_Hyst}SQUID measurements ($m-H$-loops) for different YIG/GIG bilayers measured at a temperature of 100\,K with a maximum applied field of 50\,mT.}
\end{figure}

To characterize the magnetic properties of the heterostructures, the magnetization vs field ($m-H$) dependence was measured with a magnetic field applied within the sample plane. The $m-H$ of YIG/GIG for magnetic fields up to 50\,mT are obtained by measuring the GGG/YIG/GIG sample and subtracting a linear fit to compensate for the paramagnetic contribution of the substrate.
Fig.\ \ref{fig:SQUID_Hyst} (a) shows the $m-H$ curves for single layer GIG and YIG measured at a temperature of $T=100$\,K. 
Note that the magnetic moment of the YIG layer of around $0.6\times10^{-7}$\,Am$^2$ corresponds to a magnetization of 133\,kA/m, which is similar to other YIG thin films and bulk samples \cite{Onbasli2014}, confirming the high quality of the thin films. Since heterostructures of varying thickness are investigated, in the following we will focus on the total magnetic moment of the layers rather than on the magnetization.
Fig.\ \ref{fig:SQUID_Hyst}(b-d) show the $m-H$ curves measured at the same temperature for YIG/GIG heterostructures of varying YIG thickness. All samples generally show sharp switching features, however, the YIG(36\,nm)/GIG(30\,nm) (Fig.\ \ref{fig:SQUID_Hyst} (d)) shows segmented switching features in this magnetic field range. The curve displays an ‘inner hysteresis’ and additionally a larger hard axis loop. Secondly, the total magnetization of the bilayers at $\mu_0H = 20$\,mT decreases with increasing YIG layer thickness. This behavior indicates that the net moments of YIG and GIG are antiparallel in the low-field region. 

\begin{figure}
	\includegraphics[width=8.5cm]{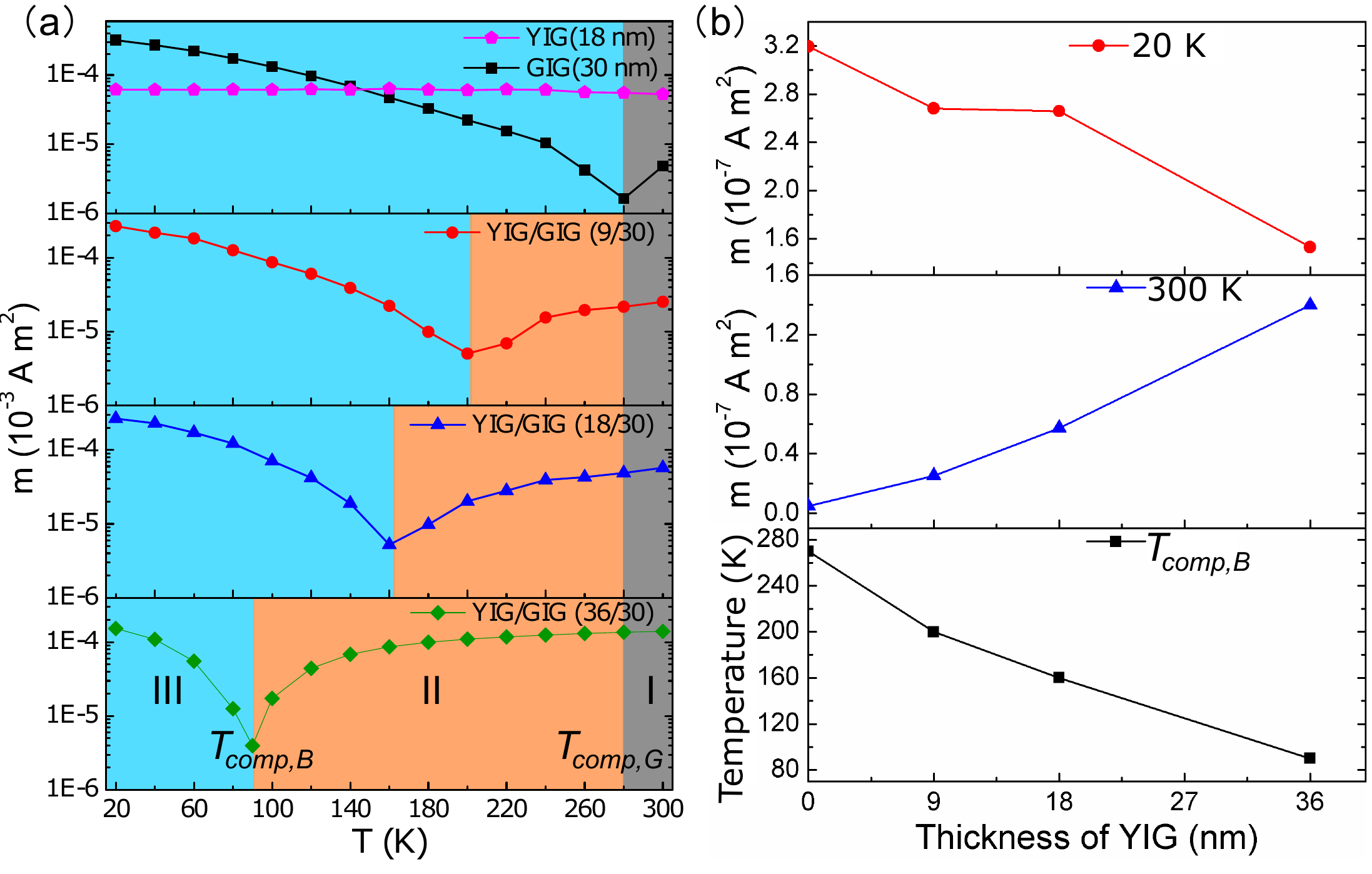}
	\caption{\label{fig:SQUID_LF}(a) $M-T$ curves of the YIG, GIG and YIG/GIG films in low magnetic field measured by $m-H$-loops at different temperatures. The shaded regions indicate $T>\TcompG$ (zone I), $\TcompG> T>\TcompB$ (zone II), $T<\TcompB$ (zone III) for the respective samples. (b) The dependence of the magnetic moment $m$(300\,K), $m$(20\,K) and the bilayer compensation temperature on the YIG thickness.}
\end{figure}

In order to further understand this behavior, the temperature dependence of the $m-H$ curves was measured. The net magnetic moments $m$ of the YIG/GIG samples at 50\,mT were obtained. 
As shown in the top of Fig.\ \ref{fig:SQUID_LF} (a), the magnetic moment of the pure YIG layer is only weakly dependent on temperature, while the pure GIG sample is ferrimagnetic with a compensation temperature (\TcompG) of 280\,K, which is close to the bulk value (295\,K \cite{Pauthenet1958}). The GIG magnetization is strongly temperature dependent due to the Gd moment increasing towards lower temperatures. Above \TcompG, the direction of the magnetization of the GIG sample is given by the direction of the $d$-Fe moments, while below \TcompG, the magnetization direction is given by the direction of the $c$-Gd moments. 
Note that the antiferromagnetic coupling between the $a$-Fe and $d$-Fe sublattices within one layer cannot be broken by magnetic fields accessible in our labs. We therefore simplify the description of the magnetic properties by introducing one magnetic Fe lattice for YIG and GIG as the sum of minority $a$-Fe and majority $d$-Fe, respectively.

Having established the compensation temperature of pure GIG, we observe that when grown in a bilayer with YIG, the compensation temperature shifts to lower temperatures with increasing YIG thickness. We indicate in Fig.\ \ref{fig:SQUID_LF} three regions of the bilayer, corresponding to: above the compensation of the pure GIG film (grey, zone I), temperatures below the compensation of the bilayer (blue, zone III), and an intermediate region (orange, zone II). The shifting of the compensation temperature of the bilayer \TcompB\ indicates that there is a coupling between the YIG and GIG layer that is compensating for the change in the Gd orientation that occurs. 
The magnitude of the magnetic moments at a magnetic field of 20\,mT at different temperatures is summarized in Fig.\ \ref{fig:SQUID_LF} (b). At 20\,K, the total magnetic moment decreases with increasing YIG thickness, similar to the 100\,K measurements shown in Fig.\ \ref{fig:SQUID_Hyst}). At 300\,K, the total magnetization increases with increasing thickness of YIG. Taking into account the reversal of the magnetization direction of the magnetic sublattices in GIG, this indicates an interlayer ferromagnetic coupling in the whole temperature range of the Fe sublattices of YIG and GIG, i.e. considering the Fe moments and Fe-O bonds only, the bilayer has a coherent magnetic structure at low fields at all temperatures. 

\begin{figure}
	\includegraphics[width=8.5cm]{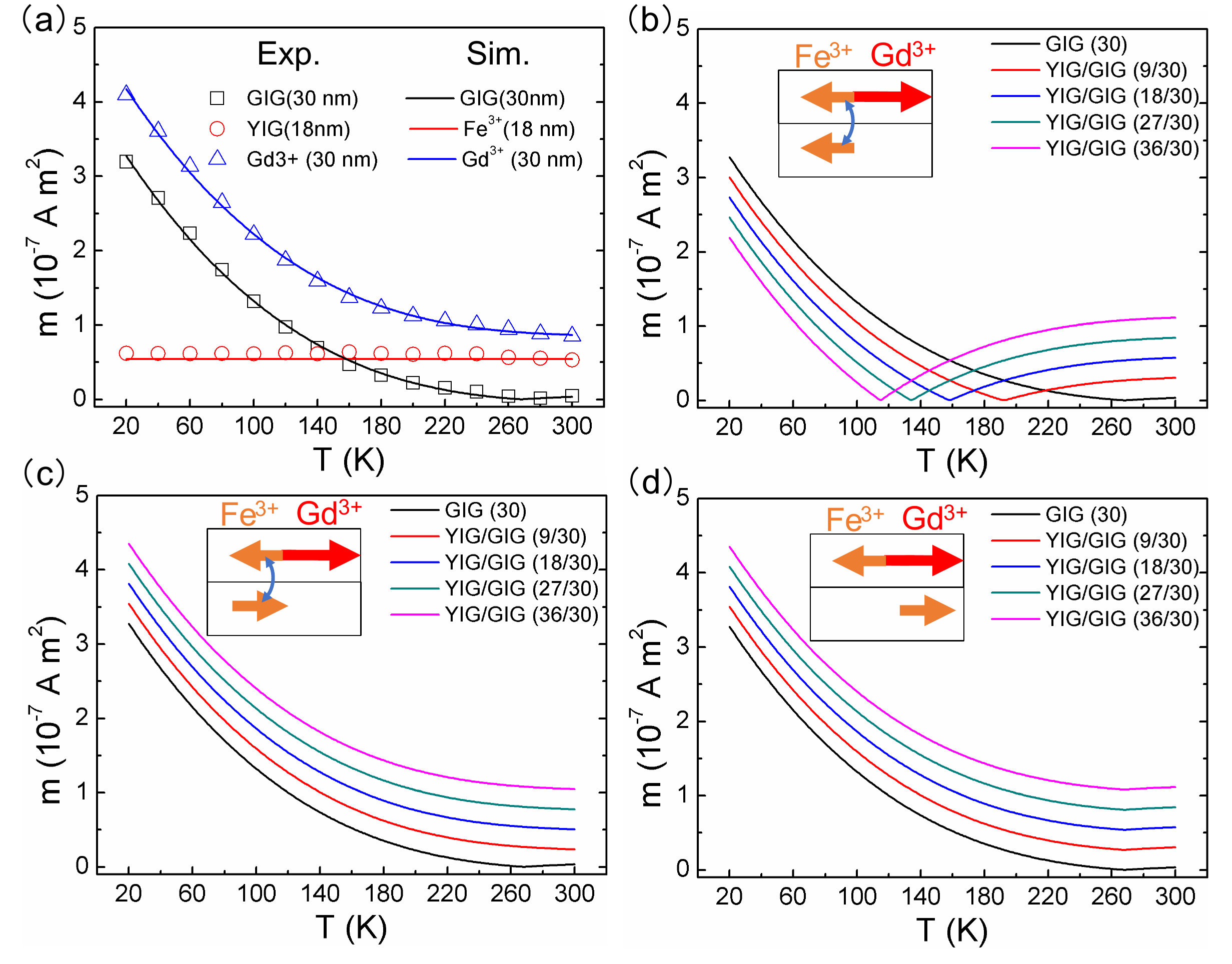}
	\caption{\label{fig:Model}(a) Modeled $m-T$ curve of the $m$(Gd) and $m$(Fe). Modeled $m-T$ curves of GIG and YIG/GIG bilayers, where $m$(Fe,GIG) of GIG and $m$(Fe,YIG) of YIG are (b) ferromagnetically coupled, (c) antiferromagnetically coupled and (d) without coupling. The insets show the magnetic sublattice alignment at low magnetic fields. The blue arrow indicate the presence of coupling.}
\end{figure}

In order to support our claim of a coherent magnetization structure of the Fe sublattices, we perform a simple simulation taking into account the temperature dependence of the Gd and Fe magnetic moments \cite{Yamagishi2005}. We model the net Fe magnetic moment as $m$(Fe) = $|m$($d$-Fe)$-m$($a$-Fe)$|$, so the magnetic moment of GIG is equal to $|m$(Fe)$-m$(Gd)$|$. Above \TcompG, the modulus $m$(Gd) = $m$(Fe)$-m$(GIG), below \TcompG, $m$(Gd) = $m$(GIG)$ + m$(Fe). Since the total magnetic moment of the Fe ions sublattices is the same in YIG and GIG, the $m$(Fe) can be seen as $m$(YIG). The magnetic moment of Gd in GIG can be extracted from the magnetization of individual YIG (18\,nm) and GIG (30\,nm) layers shown in Fig.\ \ref{fig:SQUID_LF}(a) top panel. The magnetization of YIG is only weakly temperature dependent, so we can model its magnetic moment as constant $m$(Fe,18\,nm) = $5.4\times10^{-8}$\,Am$^2$. We note that the YIG magnetization should follow $T^{3/2}$ from the Curie temperature $T_C$ to 0\,K, however, we stay far below $T_C$. 
The $m$(Gd) can be extracted from the temperature dependent magnetization curves and the data points are fitted by a third order polynomial used for the simulations. More precise modeling of $m$(Gd) can be done in principle using a self consistent molecular field acting on the Gd spin moments as input for the Brillouin function \cite{Lister1966}. This would require knowledge of the detailed temperature dependence of the Fe sublattice magnetization.
However, the exact behavior of the sublattice magnetization is out of scope of this work and the parameters from the bulk cannot be transferred to the thin films that usually have somewhat reduced Curie temperatures. Therefore, we describe $m$(Gd) only phenomenologically in the temperature range of our measurements to facilitate the analysis. The temperature dependencies of the magnetic moments of Gd$^{3+}$, Fe$^{3+}$ and simulated GIG are plotted in Fig.\ \ref{fig:Model}(a).

We model three different cases for the interlayer coupling of YIG/GIG bilayers: the respective magnetic moments $m$(Fe,GIG) of GIG and $m$(Fe,YIG) of YIG are ferromagnetically coupled, antiferromagnetically coupled and without coupling. 
For the ferromagnetic Fe-Fe coupling case, the overall magnetic moment of the system is given as $m=|m$(Fe,GIG)$+m$(Fe,YIG)$-m$(Gd,GIG)$|$; For the antiferromagnetic Fe-Fe coupling, $m=|m$(Fe,GIG)$-m$(Fe,YIG)$-m$(Gd,GIG)$|$; Without coupling $m = | m$(Fe,GIG)$ - m$(Gd,GIG)$| + m$(Fe,YIG). The respective results are plotted in Fig.\ \ref{fig:Model}. 
As shown in Fig.\ \ref{fig:Model}(b), for ferromagnetic coupling, \TcompB\ is decreasing with increasing thickness of YIG. In addition, the total magnetic moment increases with the thickness of YIG at 300\,K and the total magnetic moment decreases with the thickness of YIG at 20\,K. These features are consistent with the experimental results and support our claim that the layers are indeed ferromagnetically coupled with respect to iron atoms. For an assumed antiferromagnetic Fe-Fe coupling at the YIG-GIG interface (Fig.\ \ref{fig:Model} (c)), the bilayers do not possess compensation temperatures below 300\,K for the simulated thicknesses. Without coupling between the YIG and GIG magnetization (Fig.\ \ref{fig:Model} (d)), the bilayers never have a compensation point, but only a total magnetization minimum at \TcompG.

\begin{figure}
	\includegraphics[width=8.5cm]{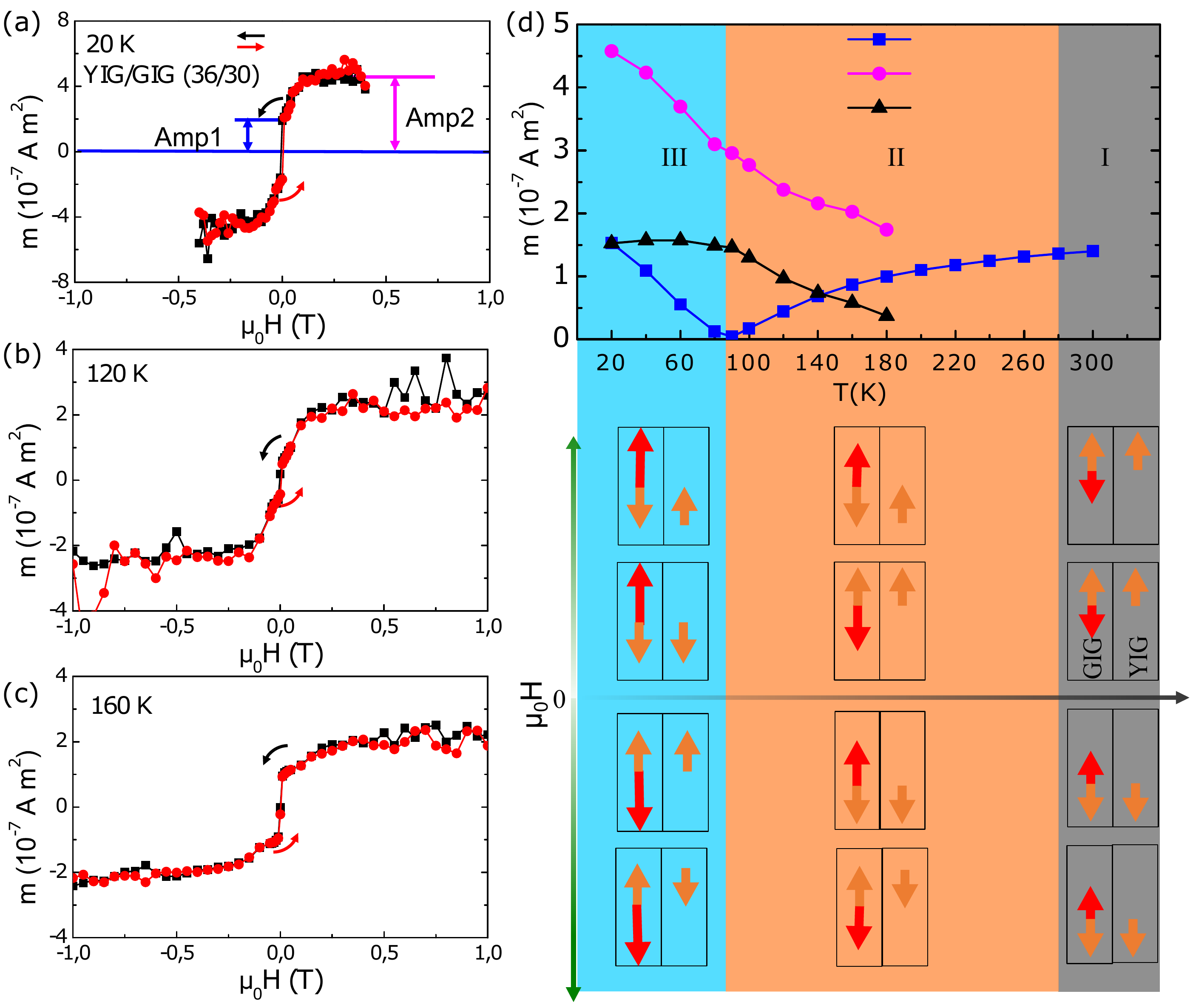}
	\caption{\label{fig:SQUID_HF}$m-H$ curves of a YIG(36)/GIG(30) sample measured at a temperature (a) 20\,K (b) 120\,K and (b) 160\,K in high magnetic  field. (d) $m-T$ curve of YIG(36)/GIG(30) with parallel state and antiparallel state and the alignment of the YIG and GIG layer in high and low magnetic fields in different temperature regions}
\end{figure}

The above discussion was on magnetization measurements at low magnetic field and we used simulations based on the individual materials to compare to the measured SQUID data. We have seen that these investigations show a ferromagnetic Fe-Fe coupling across the YIG-GIG interface. The double switching in the YIG(36nm)/GIG(30nm) sample (Fig.\ \ref{fig:SQUID_Hyst}(d)) indicates that a sufficiently large magnetic field can break the interlayer coupling. To investigate this further, we perform $m-H$ measurements exploiting larger magnetic fields at various temperatures. These $m-H$ of YIG/GIG can be obtained via measuring the entire GGG/YIG/GIG sample and then subtracting the paramagnetic contribution of GGG determined in a separate measurement. As shown in Figs.\ \ref{fig:SQUID_HF} (a)-(c), we observe double switching over a large temperature range. Further increase of the external field leads to a saturation of the overall magnetic moment. To describe the curves, we introduce $Amp2$ as the maximum amplitude of the hysteresis. Here, the Fe-Fe interlayer coupling is broken and the net moments of YIG and GIG layers align parallel so that $m$(YIG)$+m$(GIG)$=Amp2$. Approaching the compensation temperature of the GIG layer \TcompG, the signal from the GIG layer becomes too weak to perform this type of analysis with respect to the signal resolution of the SQUID and in the presence of the strong substrate background. Therefore, we evaluate $Amp2$ only up to T = 180\,K.
The ‘inner hysteresis’, as already seen in Fig.\ \ref{fig:SQUID_Hyst} (d) and in Fig.\ \ref{fig:SQUID_HF} (a), which indicates a switching at low magnetic fields has the amplitude $Amp1$. In this low-field region, the Fe-Fe coupling is not broken and the magnetic moments of the whole bilayer stack reverses by reversing the small magnetic field keeping the relative orientation of the Fe and Gd moments intact. The amplitude of this hysteresis is given by the difference of the net magnetic moments of the individual layers $|m$(YIG)$-m$(GIG)$|=Amp1$. The preservation of the relative orientation of the individual moments implies that the orientation of the net magnetic moments of the YIG and GIG layers changes going from zone II to zone III in Fig.\ \ref{fig:SQUID_LF}. In zone II, where the Gd moment is smaller than the net Fe moment of the coupled layers, the YIG layer aligns parallel to the field while the GIG layer aligns antiparallel to the field. If the magnitude of the Gd moment overcomes that of the net bilayer Fe moment in zone III, GIG aligns parallel to the field and YIG antiparallel. This is illustrated in Fig.\ \ref{fig:SQUID_HF}(d). The relative alignment of Fe (orange) and Gd (red) moments is depicted. 
In the high temperature region above the compensation temperature of the GIG layer (grey sketched zone I in Fig \ref{fig:SQUID_HF}(d)) the magnetization of GIG is dominated by the net iron moment. Here the magnetization of the YIG layer and GIG layer are always parallel to the external field at low and high field. 

We have seen in SQUID measurements that the magnetization of the YIG and GIG layers aligns differently depending on the temperature and relative thickness. Tuning the thickness of the respective layers allows for choosing the magnetic properties of the heterostructures. The larger the YIG:GIG ratio, the lower the bilayer compensation temperature \TcompB. We can thereby functionalize the coupled layers and build devices that have defined relative orientation of the magnetic moments.
However, SQUID measurements do not distinguish which layer switches (top or bottom). In order to disentangle this, we conduct surface-sensitive methods to probe the top surface layer individually to then, in combination with SQUID, identify which layer switches at which field.

\begin{figure}
	\includegraphics[width=8.5cm]{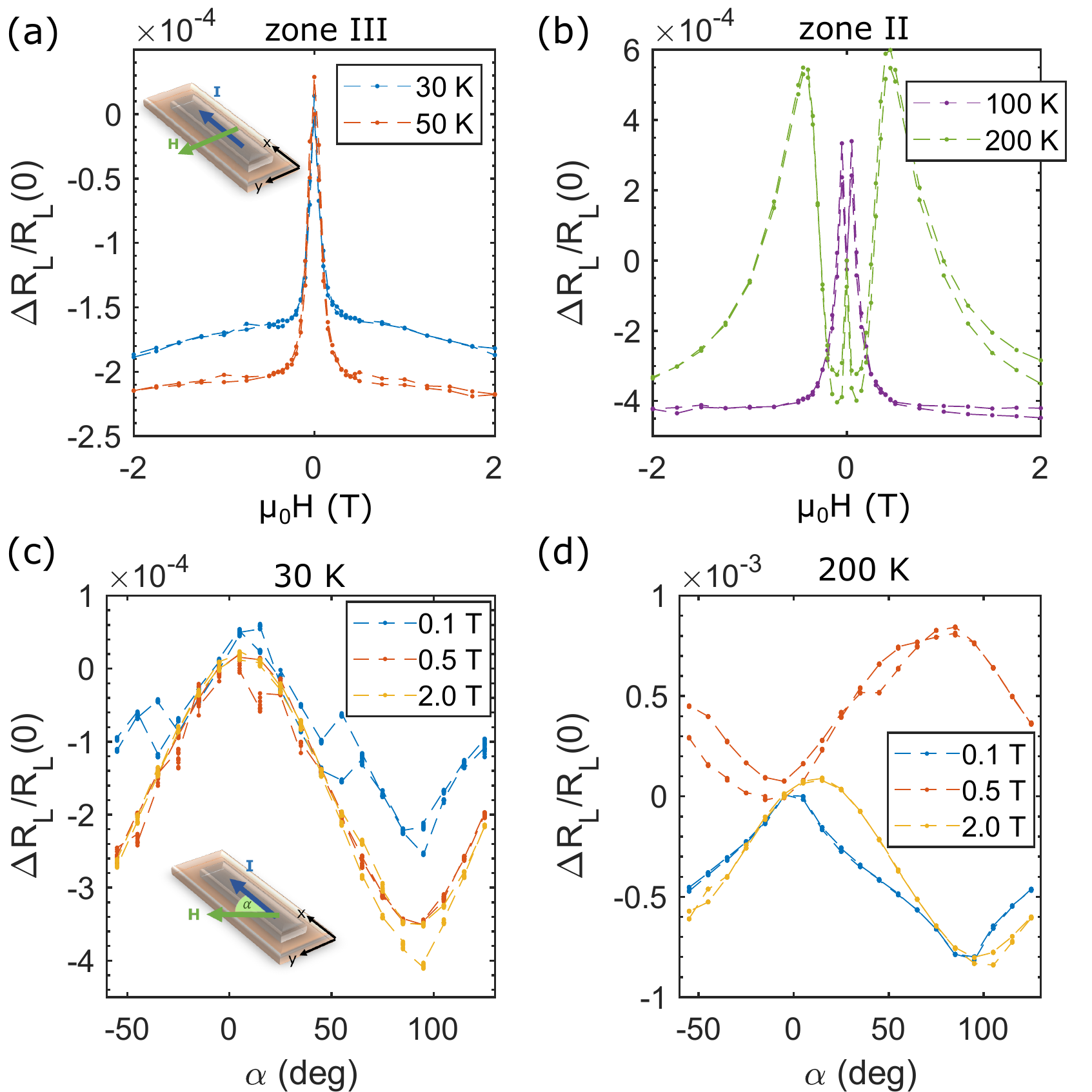}
	\caption{\label{fig:SMR}Uniaxial SMR measurements at a GGG/YIG(36)/GIG(30)/Pt sample at various temperatures (a-b) with the field applied in the sample plane perpendicular to the current as depicted in the inset of (a). ADMR for two different temperatures (c-d) where the field is rotated within the sample plane as illustrated in the inset of (c).}
\end{figure}

A suitable tool to investigate the magnetic properties of the top layer is spin Hall magnetoresistance (SMR) \cite{Nakayama2013,Chen2016}. The spin accumulation in a heavy metal in close contact with a ferrimagnetic insulator interacts with the magnetic order parameter only at the very interface. The resistance of a Pt bar defined on the surface of a GGG/YIG(36)/GIG(30) sample is modulated by $\Delta R_L\propto (1-m_y^2)$ \cite{Chen2016}, where $m_y$ is the magnetization component of the GIG layer in plane perpendicular to the Pt bar. We normalize the change of resistance to the zero-field value.
We perform uniaxial measurements with the magnetic field applied in-plane perpendicular to the Pt bar. The uniaxial field measurements at 30\,K and 50\,K describe a sharp drop of resistance at low magnetic fields followed by negligible further resistance changes, which is shown in Fig.\ \ref{fig:SMR} (a). The sharp decrease of resistance is likely due to the annihilation of differently aligned domains, going from a multidomain state to a monodomain state when the GIG net magnetic moment aligns with the field. In the absence of an external field, the sample symmetry allows for domains with magnetization aligned along the Pt bar. 
At temperatures above \TcompB, double switching is observed in the SMR data with increasing magnetic field (see Fig.\ \ref{fig:SMR} (b)). This indicates the rotation of the GIG magnetic moment after the interfacial coupling between YIG and GIG is broken. The continuous increase of resistance in the switching process indicates that the top layer switching is in fact not an abrupt process, but a gradual rotation since only the alignment of the GIG magnetization away from the y-direction can increase the resistance due to SMR. We confirm this by performing rotation measurements, where the field rotates within the sample plane ($\alpha$-plane) from around $-55^\circ$ to $125^\circ$, relative to the current direction, and back. The experiment is performed at 30\,K, which is in zone III (Fig.\ \ref{fig:SMR} (c)) and at 200\,K, which is in zone II (Fig.\ \ref{fig:SMR} (d)). The longitudinal resistance is given as relative to the resistance value at 0° (along the Pt bar). At low temperatures, where no top-layer switching is observed in the uniaxial SMR measurements, we also do not observe any changes in the phase of the angular-dependent magnetoresistance (ADMR) for magnetic fields of different magnitude. However, increasing the temperature to 200\,K, the ADMR becomes strongly field-dependent. At low magnetic field and at high magnetic field, the phase of the ADMR is close to $0^\circ$, indicating the alignment of the GIG magnetization axis with the external magnetic field. At 0.5\,T, however, in the region of the GIG switching, a phase shift of almost $90^\circ$ is observed, indicating the perpendicular alignment of GIG magnetization to the magnetic field in the switching region. 

\begin{figure} 	
	\includegraphics[width=8.5cm]{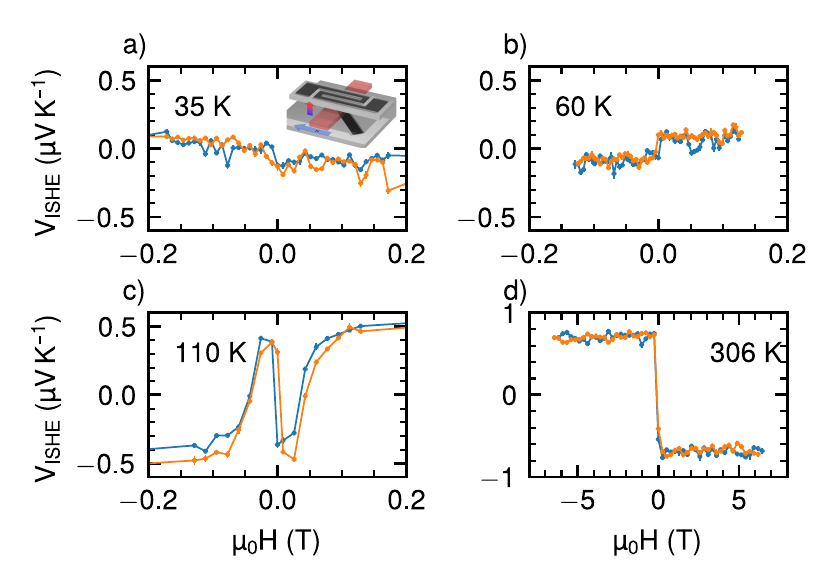} 
	\caption{\label{fig:SSE}$V_{ISHE}-H$ loops of a GGG/YIG(36)/GIG(30)/Pt sample at various temperatures (a) 35\,K, (b) 60\,K, (c) 110\,K and (d) 306\,K. For (d) a larger field range is displayed. $V_{ISHE}$ is normalized by the estimated temperature gradient and a constant offset of the signal is subtracted.} 
\end{figure}

A complementing tool to investigate the magnetic properties of the bilayer system is the longitudinal spin Seebeck effect (SSE). The SSE can be used to investigate the magnetic properties close to \TcompG\ at which SQUID measurements cannot resolve the magnetization of the GIG layer.
For the SSE measurement, the sample is exposed to an out-of-plane temperature gradient by sandwiching it between a temperature sensor and a resistive heater (see inset of Fig.\ \ref{fig:SSE} (a)). The temperature gradient is estimated by monitoring the resistance of sensor and heater. The external magnetic field is applied in the sample plane, perpendicular to the temperature gradient. 
Measuring the thermal excitation of spin waves in such a bilayer can potentially lead to superposing spin currents originating from YIG and GIG, measured as a voltage $V_{ISHE}$ via the inverse spin Hall effect (ISHE) in a heavy metal top layer \cite{Saitoh2006, Cramer2017}. The SSE of YIG(36)/GIG(30)/Pt(4) was measured at different temperatures. As shown in Fig.\ \ref{fig:SSE}, a hysteretic voltage signal $V_{ISHE}$ is obtained by sweeping the magnetic field.  
At the lowest temperature of 35\,K, we observe a negative amplitude of the SSE as shown in Fig.\ \ref{fig:SSE}(a). Going to 60\,K, the sign of the measured voltage changes its sign to positive (Fig.\ \ref{fig:SSE}(b)). For 110\,K, the shape of the SSE signal fundamentally changes as seen in Fig.\ \ref{fig:SSE}(c). At low magnetic fields, a negative switching is observed before the signal changes sign again at around 40\,mT. At 306\,K, only one switching event is visible at low magnetic fields, having a negative sign (Fig.\ \ref{fig:SSE}(d)). 

These features emphasize the SSE signal to have it's origin in the GIG layer. We investigate these features in the following, by taking into account the magnetic properties of the bilayer system determined by SQUID and the complex behavior of SSE signal measured at pure GIG samples \cite{Gepraegs2016,Cramer2017}. 
For temperatures below the compensation temperature of the bilayer \TcompB, we can compare the SSE measurement with the $m-H$ loop. In the SQUID measurement at 60\,K (zone III), a switching of YIG is expected at approximately 30\,mT. However, we do not observe this feature at 30\,mT in SSE measurements. Thus, a possible contribution to the spin current originating from YIG seems to be damped in the GIG layer in this sample. At temperatures above \TcompB\ and below \TcompG\ (zone II), a segmented switching is observed. This is expected for the GIG sensitive SSE measurement, as the GIG layer switches at higher fields in this temperature range (thus the second step in the field sweep). 
The sign change of the SSE signal $V_{ISHE}$ at low temperatures also suggests the GIG layer to be the main spin current source. For even lower temperatures ($T = 35$\,K), the SSE signal undergoes a sign change, which can be explained by the change of spectral weight and temperature dependence of occupied magnon modes in GIG \cite{Cramer2017, Gepraegs2016}. For pure GIG, another sign change is observed at compensation temperature \TcompG, because of the reversal of the sublattice magnetization for a constant external applied magnetic field \cite{Cramer2017}.  
As the SSE signal follows the GIG magnetization, we can investigate the orientation of the GIG at higher temperatures. When changing the temperature from below to above \TcompG, there is no sign change observed in the YIG/GIG bilayer. For a coupling of YIG and GIG via the net moment of each film, a reversal of the sign of $V_{ISHE}$ would be expected at \TcompG, which is not supported by our data. This is in line with our claim, that the Fe sublattice moments of YIG and GIG dominate the coupling, thus when changing the temperature across \TcompG\ the GIG orientation stays the same and no reversal of the sign of $V_{ISHE}$ is observed. 

\section{Discussion}
Together, our measurements indicate that we have robust Fe-Fe exchange coupling between defined YIG and GIG layers. While SQUID gives a basis of the interpretation of the magnetic behavior of the heterostructures, both SSE and SMR selectively show the behavior of the top GIG layer only. Our high quality thin films access a phase space of samples with arbitrarily aligned magnetization of the bilayers, depending of the relative thicknesses. Up to now, the exchange coupling between YIG and GIG has only been described as an interface effect in GGG/YIG samples \cite{Gomez2018, Li2020, Kumar2021}. The artificial stacks grown here give much better control over the bilayer properties and allow for fine tuning of these by choosing the relative thickness of YIG and GIG. 
We identify three temperature regions, where the samples have fundamentally different responses to an external magnetic field. 
Above the GIG compensation point \TcompG, the coupled net iron moments are always parallel and show in the direction of the external magnetic field. Reducing the temperature below \TcompG, we still observe ferromagnetic Fe-Fe coupling across the YIG/GIG interface at low magnetic fields. Increasing the field, we break this coupling across the interface and the Gd moment of the GIG layer orients with the external field. Uniaxial SMR measurements show the switching of the top GIG layer and rotation measurements indicate a continuous rotation of the magnetic moments in the reorientation regime. Decreasing the temperature further, we observe a second compensation point, where the net GIG layer moment equals the YIG layer moment. We label this bilayer compensation point \TcompB. We show that \TcompB\ can be tuned by choosing the relative thickness of the YIG and GIG layers. This allows for devices with controllable magnetization direction. Below \TcompB, we observe again a double switching in SQUID, indicating the reorientation of one of the layers. Both SMR and SSE measurements do not show this switching, indicating that the bottom YIG moment rotates, leaving the top GIG moment aligned with the field. 
SSE measurements thereby prove to be sensitive to the top GIG layer only, since the switching of the bottom YIG spin current source is not observed in our measurements. Moreover, SSE measurements show a change of sign at certain temperature, which is not associated with the direction of the net magnetic moment, but with the magnon population as observed in pure GIG layers \cite{Gepraegs2016, Cramer2017}. Recently, it was reported that magnon hybridization may occur in YIG-GIG heterostructures leading to a reduction of the temperature of the sign change. 
We note that here, the change of sign occurs at around 35\,K for a YIG(36\,nm)/GIG(30\,nm) sample. SSE measurements performed at pure GIG layers have shown the sign change to occur at around 44\,K to 72\,K \cite{Cramer2017}, depending on several parameters like magnetic compensation point, heavy metal material and GIG-heavy metal interface quality. The here observed sign change temperature of 35\,K is remarkably low compared to these reports, which might indicate magnon hybridization. However, a direct comparison with literature values is difficult due to the strong dependence on the surface quality of the samples. A simultaneous deposition of the HM layer excludes a different GIG/HM interface for different samples and will be target of futures studies. \\
For the interlayer coupling strength between YIG and GIG, we assume that the coupling energy is equal to the gain of Zeeman energy to align both YIG and GIG magnetization in zones II and III. We calculate the energy at the low temperature state of sample YIG(36\,nm)/GIG(30\,nm). From Fig.\ 5 (a) we extract the saturation field at 20\,K, which is the field at which $Amp2$ is reached as $\mu_0H = 0.1$\,T. Rotating $m$(YIG) from antiparallel to parallel leads to a gain in Zeeman energy by $2\cdot m($YIG$) \cdot H$. $m$(YIG) is given by $m$(YIG)= $\frac{1}{2}(Amp2-Amp1) \approx 1.5\times10^{-7}\,$Am$^2$. Taking into account the interface area of 0.25\,cm$^2$, we estimate an effective interface coupling energy of 0.0012\,J/m$^2$. With 8\,Fe atoms (d or a site) on the surface of a (001) oriented unit cell this relates to 1.4\,meV per Fe atom at the interface. Comparing this to the dominant exchange parameter $J_1$ in YIG, which is found to be 6.8\,meV \cite{Princep2017} and $J_1$ in GIG of 4.0\,meV  \cite{Gomez2018}, we observe a qualitative agreement, but the interlayer coupling strength appears to be smaller than the direct exchange coupling in pure YIG and pure GIG but of the same order of magnitude as the theoretical limits. For interfacial coupling of YIG-GIG at the YIG/GGG interface, a coupling strength of 0.14\,meV was discussed by Gomez et al. based on the c-Gd antiferromagnetic interaction $J_{cd}$ to d-Fe \cite{Gomez2018}. 
Our value is significantly larger than this. This can be understood by the fact that they investigate an ultra-thin layer formed by interdiffusion with the substrate. In our case, however, the $J_{cd}$ interaction energy is not a limiting factor as it needs to be integrated over the volume/thickness of the Gd layer and the resulting energy is then much larger than the interfacial exchange coupling between YIG and GIG layers that is limited by Fe-Fe interactions at the interface. 
In reality, the competition between different exchange interactions can easily lead to more complicated spin structures than in our fully collinear toy model discussed above and we see hints for this in the continuous rotation of the magnetization observed in SMR measurements. Also in our samples we have the YIG-GGG interface at the substrate and therefore the coupling effects induced by intermixing with the substrate should be present. However, as this interface is ultrathin and contributes very little to the total sample magnetization it can be neglected to first order in our analysis and it was only observed at very low temperatures by Gomez at al. \cite{Gomez2018}. Traces of this intermixing GIG at the substrate interface might be seen in the low amplitude SQUID curves shown in Fig.\ 2, where the pure YIG as well as the bilayers with a thin YIG layer appear slightly exchange-biased. In spite of the limitations of the simplified toy model for the analysis we can safely state that the largest interfacial coupling dominating the properties of our bilayers stems from the upper YIG/GIG interface.

\section{Conclusion} 
YIG/GIG bilayers were fabricated by PLD and the magnetic coupling of the samples was investigated by SQUID, SSE and SMR. It is found that the YIG/GIG bilayers show ferrimagnetic features. The compensation temperature of the bilayer system shifts to lower temperatures with increasing thickness of the YIG layer, which originates from the effective ferromagnetic coupling of the iron magnetic sublattices at the YIG-GIG interface, i.e.\ in the epitaxial bilayer, the iron atoms spin subsystem on respective d-Fe and a-Fe sites is coherent over the boundary of the two materials in zero magnetic field. Below the compensation temperature of the GIG layer, an external magnetic field can break the coupling, leading to a parallel magnetization of YIG and GIG layers at high field. At low magnetic fields, the orientation of the YIG and GIG magnetic moment depends on the temperature (Gd moment). SMR and SSE measurements reveal the behavior of the top GIG layer, from which the temperature and field dependence of alignment of YIG and GIG can be obtained. Our results demonstrate that the magnetic coupling in insulating YIG/GIG heterostructures can be manipulated analogously to that of metallic spin valves and open the pathway to manipulate the magnon transport. 

\section*{Acknowledgments}
The authors gratefully acknowledge funding by Deutsche Forschungsgemeinschaft (DFG, German Research Foundation) Project No. 358671374. This work was supported by the Max Planck Graduate Center with the Johannes Gutenberg–Universität Mainz (MPGC) as well the Graduate School of Excellence Materials Science in Mainz (GSC266). This work was funded by the Deutsche Forschungsgemeinschaft (DFG, German Research Foundation) Spin+X (A01+B02) TRR 173 - 268565370. This work is partially supported by the National Key R\&D Program of China (Grant No. 2018YFB0704100), the National Science Foundation of China (Grants No. 11974042, No. 51731003, No. 51927802, No. 51971023, No. 11574027, and No. 61674013). Sally Lord gratefully acknowledges the DAAD RISE Germany Scholarship.

\bibliography{BibTex1}

\end{document}